\documentclass[fleqn]{article}
\usepackage{amsmath,amssymb}
\usepackage{epsfig,graphicx}

\catcode`@=11 \@addtoreset{equation}{section} \catcode`@=12

\begin{document}
\title{\vskip-1.7cm \bf  Statistical sum in the CFT driven cosmology}
\date{}
\author{A.O.Barvinsky$^{1,\,2}$ and Yu.V.Gusev$^{3,\,4}$}
\maketitle \hspace{-8mm} {\,\,$^{1}$\em
Theory Department, Lebedev
Physics Institute, Leninsky Prospect 53, Moscow 119991, Russia\\
$^{2}$Department of Physics, Ludwig Maximilians University,
Theresienstrasse 37, Munich, Germany\\
$^{3}$Lebedev Research Center in Physics, Leninsky Prospect 53, Moscow 119991, Russia\\
$^{4}$IRMACS Centre, Simon Fraser University,
8888 University Drive, Burnaby, B.C. V5A 1S6 Canada}
\begin{abstract}
The path integration technique recently developed for the
statistical sum of the microcanonical ensemble in cosmology is
applied to the calculation of the one-loop preexponential factor in
the cosmological model generated by a conformal field theory with a
large number of quantum species -- the model of initial conditions
possibly related to the resolution of the cosmological constant and
landscape problems. The result is obtained for the family of
background cosmological instantons with one oscillation of the FRW
scale factor. The magnitude of the prefactor is analytically and
numerically estimated for fields of various spins conformally
coupled to gravity, which justifies the validity of semiclassical
expansion for this family of cosmological instantons.
\end{abstract}

\section{Introduction}
In this paper we apply the path integration technique recently
developed for the statistical sum of the generic
time-parametrization invariant gravitational system with the FRW
metric \cite{PIQC} to a particular case of the cosmology generated
by the conformal field theory (CFT) with a large number of quantum
species. This problem is motivated by the recently suggested model
of initial conditions in cosmology in the form of the microcanonical
density matrix \cite{slih,why}. When applied to cosmology with a
large number of fields conformally coupled to gravity, this theory
can be important within the cosmological constant and dark energy
problems. In particular, its statistical ensemble is bounded to a
finite range of values of the effective cosmological constant, it
generates an inflationary stage and is potentially capable of
generating the cosmological acceleration phenomenon within the
so-called Big Boost scenario \cite{bigboost}.

As shown in \cite{why}, for a spatially closed cosmology with
$S^3$-topology the microcanonical statistical sum can be represented
by the Euclidean quantum gravity path integral,
    \begin{eqnarray}
    &&Z=
    \!\!\int\limits_{\,\,\rm periodic}
    \!\!\!\! D[\,g_{\mu\nu},\phi\,]\;
    e^{-S[\,g_{\mu\nu},\phi\,]},
    \end{eqnarray}
over the metric $g_{\mu\nu}$ and matter fields $\phi$ which are
periodic on the Euclidean spacetime with a compactified time $\tau$
(of $S^1\times S^3$ topology). The FRW metric arises in this path
integral as the set of major collective variables of cosmology.
Under the decomposition of the full set of $g_{\mu\nu}(x),\phi(x)$
into the minisuperspace FRW sector
    \begin{eqnarray}
    ds^2 =N^2(\tau)\,d\tau^2
    +a^2(\tau)\,d^2\Omega^{(3)}, \label{FRW}
    \end{eqnarray}
and inhomogeneous ``matter'' fields
$\varPhi(x)=(\phi(x),\psi(x),A_\mu(x), h_{\mu\nu}(x),...)$ on the
background of this metric the path integral can be cast into the
form of an integral over a minisuperspace lapse function $N(\tau)$
and a scale factor $a(\tau)$,
    \begin{eqnarray}
    &&Z=\int
     D[\,a,N\,]\;
    e^{-\varGamma[\,a,\,N\,]},              \label{1}\\
    &&e^{-\varGamma[\,a,\,N]}
    =\int D\varPhi(x)\,
    e^{-S[\,a,\,N;\,\varPhi(x)\,]}\ .             \label{2}
    \end{eqnarray}
Here, $\varGamma[\,a,\,N\,]$ is the Euclidean effective action of
the fields $\varPhi$ (which include also the metric perturbations
$h_{\mu\nu}$) on the FRW background, and
$S[\,a,N;\varPhi(x)\,]\equiv S[\,g_{\mu\nu},\phi\,]$ is the
original action rewritten in terms of this minisuperspace
decomposition. It is important that this representation is not a
minisuperspace approximation, when all the fields $\varPhi(x)$ are
frozen out. Rather this is the disentangling of the collective
degrees of freedom from the configuration space, the rest of which
effectively manifest itself in terms of this effective action.

Semiclassically the integral (\ref{1}) is dominated by the contribution of the saddle point -- the periodic solution of the effective Friedmann equation of the {\em Euclidean} gravity theory
$\delta\varGamma/\delta N(\tau)=0$,\footnote{In view of the time-parametrization invariance of $\varGamma[\,a,\,N\,]$ the second equation $\delta\varGamma/\delta a(\tau)=0$ is the derivative of the Friedmann equation and should not be imposed independently of the latter \cite{PIQC}.}
    \begin{eqnarray}
    &&Z=P\,\exp\big(-\varGamma_0).      \label{Z}
    \end{eqnarray}
Here $\varGamma_0=\varGamma[\,a,N\,]$ is taken at this solution and $P$ is the preexponential factor accumulating perturbation corrections of the semiclassical expansion. For the CFT driven cosmology the instanton solutions were obtained in \cite{slih} as a function of the primordial cosmological constant, whereas the calculational technique for the one-loop prefactor $P$ was developed in \cite{PIQC}.

Here we calculate this prefactor for the set of instantons in a special range of the cosmological constant in which they have a one-fold nature -- a single oscillation of the scale factor $a(\tau)$ between its minimal and maximal values.  In Sect.2 we give the description of the full set of instanton solutions of \cite{slih} and compare them with those of \cite{HHR} -- the paper which also considers trace anomaly driven cosmology but differs from our work by setting the problem and the choice of the field-theoretical model. Then after a brief formulation of the calculational method of \cite{PIQC} in Sect.3 we calculate the one-loop prefactor $P$ in Sect.4. There we begin with the description of the above mentioned set of one-fold instantons and their
various characteristics and then estimate the magnitude of $P$ for quantum fields of various spins. The paper is accomplished with concluding remarks in Conclusions.

\section{Microcanonical instantons in the CFT driven cosmology}

In the theory with a primordial cosmological constant $\Lambda$ and a large number of free (linear) fields $\phi$ conformally coupled to gravity -- conformal field theory (CFT),
    \begin{eqnarray}
    &&S[\,g_{\mu\nu},\phi\,]=
    -\frac1{16\pi G}
    \int d^4x\,
    g^{1/2}\,\Big(R
    -2\Lambda\Big) +S_{CFT}[\,g_{\mu\nu},\phi\,],
    \end{eqnarray}
the effective action $\varGamma[\,a,N\,]$ is dominated by the
contribution of these fields because they simply outnumber the
non-conformal fields including, in particular, the graviton. Then
this quantum effective action is exactly calculable as a {\em
functional} of histories $(a(\tau),N(\tau))$ by the conformal
transformation converting (\ref{FRW}) into the static (Einstein
Universe) metric with $a={\rm const}$ \cite{FHH,Starobinsky,conf}.
The result reads \cite{slih}
    \begin{eqnarray}
    &&\varGamma[\,a,N\,]=
    \oint d\tau\,N {\cal L}(a,a')+ F(\eta),    \label{action1}\\
    &&\eta=\oint d\tau\,\frac{N}a,              \label{eta}
    \end{eqnarray}
where $a'\equiv da/Nd\tau$ and the integration runs over the period
of $\tau$ on the circle $S^1$ of $S^1\times S^3$. Here the effective
Lagrangian of its local part ${\cal L}(a,a')$ includes the classical
Einstein term with the renormalized Planck mass
$m_P=(3\pi/4G)^{1/2}$ and the cosmological constant $\Lambda=3H^2$
and contains also the contribution of the conformal anomaly of
quantum fields and their vacuum (Casimir) energy,
    \begin{eqnarray}
    &&{\cal L}(a,a')=m_P^2\left\{-aa'^2
    -a+ H^2 a^3+B\left(\frac{a'^2}{a}
    -\frac{a'^4}{6 a}+\frac1{2a}\right)\,\right\}, \label{effaction}\\
    &&F(\eta)=\pm\sum_{\omega}
    \ln\big(1\mp e^{-\omega\eta}\big).
    \end{eqnarray}
The constant $B>0$ is determined by the coefficient of the
Gauss-Bonnet term in the overall conformal anomaly of all CFT fields
$\phi$ \cite{Duffanomaly}. A nonlocal part of the action $F(\eta)$
is the free energy of their quasi-equilibrium excitations with the
temperature given by the inverse of the conformal time (\ref{eta}).
This is a typical boson or fermion sum over field oscillators with
energies $\omega$ on a unit 3-sphere.

Semiclassically the integral (\ref{1}) is dominated by the saddle
points --- solutions of the effective Friedmann equation of the {\em
Euclidean} gravity theory
    \begin{eqnarray}
    &&-\frac{a'^2}{a^2}+\frac{1}{a^2}
    -B \left(\,\frac{a'^4}{2a^4}
    -\frac{a'^2}{a^4}\right) =
    \frac\Lambda3+\frac{C}{ a^4},          \label{efeq}\\
    &&C = \frac{B}2+\frac1{m_P^2}
    \frac{dF(\eta)}{d\eta},                \label{bootstrap}
    \end{eqnarray}
modified by the quantum $B$-term and the radiation term $C/a^4$. The
constant $C$ here characterizes the sum of the renormalized Casimir
energy $B/2$ and the energy of the gas of thermally excited
particles
    \begin{eqnarray}
    \frac{dF(\eta)}{d\eta}=
    \sum_\omega\frac{\omega}{e^{\omega\eta}\mp 1}. \label{energy}
    \end{eqnarray}
Thus, the amount of thermal radiation in this self-consistent back
reaction problem is not put by hands, but determined by the {\em
bootstrap} equation (\ref{bootstrap}) which yields $C$ as a
functional of the FRW geometry in which this radiation evolves.

The inverse temperature $\eta$ -- the instanton period in units of
the conformal time -- is given by the integral (\ref{eta}) over the full period of $\tau$ or the $2k$-multiple of the integral between the two neighboring turning points of the scale factor history $a(\tau)$, $a'(\tau_\pm)=0$. This $k$-fold nature implies that in the background solution during the full period of its Euclidean time the scale factor oscillates $k$ times between its
maximum and minimum values $a_\pm=a(\tau_\pm)$, $a_-\leq a(\tau)\leq
a_+$,
    \begin{eqnarray}
    a^2_\pm=
    \frac1{2H^2}\big(\,1\pm\sqrt{1-4H^2{C}}\,\big), \label{apm}
    \end{eqnarray}
and forms a kind of a garland of $S^1\times S^3$ topology with
oscillating $S^3$ sections. As shown in \cite{slih}, these
garland-type instantons exist only in the limited range of the
cosmological constant $\Lambda=3H^2$, $0<H^2_{\rm min}<H^2< H^2_{\rm
max}=1/2B$ and are weighted in the relevant statistical ensemble by
their exponentiated on-shell action according to (\ref{Z}), where
$\varGamma_0=\varGamma[\,a,N\,]$ is taken at the solution of
Eqs.(\ref{efeq})-(\ref{bootstrap}). In particular, a set of instantons with $k=1$ -- a single oscillation of the scale factor $a(\tau)$ between its minimal and maximal values (\ref{apm}) --
occupies a certain continuous range of $H^2$ inside
$[H^2_{\rm min},\,H^2_{\rm max}]$ adjacent to its lower boundary \cite{slih}.

The main effect of these exponential weights is a complete suppression of another set of solutions -- the vacuum Hartle-Hawking instantons \cite{noboundary} with no radiation $dF/d\eta=0$, which represent the Euclidean de Sitter spacetime. These purely vacuum contributions are ruled out by their infinite {\em positive} effective action (cf. $1/a$-factor in the kinetic $B$-terms of the effective action (\ref{effaction}) which render its integrand diverging to $+\infty$ at $\tau_-$ with $a\to 0$ and $a'\to 1$). Otherwise, these exponential weights are of the same order of magnitude and they are inefficient to select most probable configurations. This makes important the calculation of the one-loop prefactor $P$ which is the main focus of this paper. The technique for this prefactor was developed for a parametrization invariant system (\ref{action1}) with a generic Lagrangian ${\cal L}(a,a')$ restricted only by the order of derivatives of $a$.\footnote{Note that the local Lagrangian of (\ref{effaction}) does not contain higher order derivatives of $a$. This is the result of the special renormalization \cite{slih}, which does not introduce into the minisuperspace sector of Einstein theory extra degrees of freedom. This choice was motivated in \cite{slih} by ghost-free requirements and certain universality properties which, in particular, relate the value of the Casimir energy in (\ref{effaction}), $B/2a$, to the coefficient $B$ of the conformal anomaly \cite{universality,DGP/CFT}.} Below we apply this technique to the CFT diven cosmology of the above type for a set of one-fold instantons with $k=1$.

Cosmological instantons in the trace anomaly driven cosmology were also considered in \cite{HHR}. It should be emphasized, however, that setting the problem and the field-theoretical model itself in this work are essentially different from our approach in \cite{slih}. To begin with, the solutions of \cite{HHR} correspond to the Hartle-Hawking no-boundary wavefunction and have the $S^4$-topology in contrast to our periodic $S^1\times S^3$ instantons associated with the microcanonical statistical sum. Therefore, the amount of radiation in \cite{HHR} is a freely specifiable constant of integration, while here it is strictly fixed by the ``bootstrap" equation (\ref{bootstrap}). Secondly, the freedom of UV renormalization is not used in \cite{HHR} to eradicate higher order derivatives from the Friedmann equation, whereas in \cite{slih} this freedom is fixed by the requirement to preserve the classical number of physical degrees of freedom and the stability of the theory against higher-derivative ghosts (cf. footnote 2 above). Finally, a primordial (or renormalized) cosmological constant is assumed to be zero in \cite{HHR}, whereas here $\Lambda$ belongs to a nontrivial range where it parameterizes the istanton solutions.

As a result the so-called double-bubble instantons of \cite{HHR} do not exist in our setting. This automatically removes the collapsing universes found in \cite{HHR} by analytic continuation to the physical Lorentzian time. Instead we have a thermal version of the Hartle-Hawking instantons -- the instanton garlands of \cite{slih} which give rise to expanding Lorentzian universes with different values of the primordial cosmological constant $\Lambda$. As discussed in \cite{bigboost}, they contain a rapidly diluting radiation of conformal particles and have a finite inflationary stage terminating with the decay of $\Lambda$ if the latter has a composite nature of a slowly varying inflaton field. This justifies the use these instantons as initial conditions for the Universe, capable of formation of the observable large scale structure.

Note that in our setting even the vacuum Hartle-Hawking instantons  mentioned above originate from the periodic boundary conditions as a limiting case when a torus $S^3\times S^1$ gets ripped at the vanishing value of $a_-=0$ and topologically becomes a 4-sphere $S^4$. Thus, these round sphere instantons exist just like in \cite{HHR},\footnote{This follows from the fact that the effective Friedmann equation (\ref{efeq}) for $H^2=0$ exactly coincides with its analogue (2.18) in \cite{HHR} under a special choice of the UV counter term $\sim R^2$ which eradicates higher order derivatives from this equation. But, as mentioned in \cite{HHR}, the inclusion of higher derivatives induced by $R^2$ does not destroy a round sphere instanton, while adding a nonvanishing cosmological constant only modifies its effective Hubble parameter.} but as mentioned above they are completely ruled out by their infinite positive Euclidean action $\varGamma_0=+\infty$ \cite{slih} -- the diverging contribution of the conformal anomaly. Quite interestingly, this property was not observed in \cite{HHR} because the action $\varGamma_0$ was not calculated there.

\section{The one-loop prefactor for a generic time-parametrization
invariant model}

The one-loop preexponential factor of the statistical sum (\ref{Z})
was obtained in \cite{PIQC} for a one-fold instaton background. For
a theory with the generic effective action (\ref{action1}) it reads
as
    \begin{eqnarray}
    P={\rm const}\times\frac1{\sqrt{\,\displaystyle
    \left|\,1-{\mbox{\boldmath$I$}}\,
    \frac{d^2F}{d\eta^2}\right|\,}}    \label{prefactor0}
    \end{eqnarray}
where $-d^2F/d\eta^2$ is a kind of a specific heat of the thermal
bath of matter quanta and $\mbox{\boldmath$I$}$ is a special
functional of the background solution $(a(\tau),N(\tau))$. As shown
in \cite{PIQC,det} this quantity also generates the functional
determinant
    \begin{eqnarray}
    {\rm Det_*}\,{\mbox{\boldmath$F$}}
    ={\rm const}\times{\mbox{\boldmath$I$}}   \label{DetFstar}
    \end{eqnarray}
of the operator $\mbox{\boldmath${F}$}$ of small disturbances of the
canonically normalized mode of the scale factor perturbation
$\varphi=\sqrt{|{\cal D}|}\,\delta a$ propagating on this
background\footnote{Star in the notation ${\rm Det_*}$ implies
gauging out the zero mode of this operator, originating from the
residual conformal Killing symmetry of the action for $\varphi$
\cite{PIQC,det}.},
    \begin{eqnarray}
    &&\mbox{\boldmath${F}$}
    =-\frac{d^2}{d\tau^2}+\frac{g''}g,      \label{operator}\\
    &&g=a'a\sqrt{|{\cal D}|},\,\,\,\,
    {\cal D}=\frac{\partial^2\cal
    L}{\partial a'\partial a'}.            \label{g}
    \end{eqnarray}
This operator is expressed via the function $g=g(\tau)$ which
simultaneously serves as its zero mode,
$\mbox{\boldmath${F}$}g(\tau)=0$, and in its turn expresses in terms
of the time derivative of the scale factor $a'$ and the Hessian of
the Lagrangian ${\cal L}(a,a')$. Thus all the quantities are
functionals of this zero mode which itself is easily calculable by
Eq.(\ref{g}) for a given instanton background.

The quantity $\mbox{\boldmath${I}$}$ expresses in quadratures as a
functional of $g(\tau)$ as follows. For a one-fold instanton the
function $g(\tau)\varpropto a'(\tau)$ has one oscillation in the
full range of the Euclidean time forming a circle. Therefore, it has
two zeroes at the antipodal points of this circle $\tau_\pm$
corresponding to a maximum and minimum values of the scale factor.
These points mark the boundaries of the half period of the total
time range, $T=2(\tau_+-\tau_-)$, at which the periodic function
$g(\tau)$ has these two first degree zeros
    \begin{eqnarray}
    &&g(\tau_\pm)=0,\,\,\,g'(\tau_\pm)
    \equiv g'_\pm\neq 0.                    \label{conditionsong}
    \end{eqnarray}
Then $\mbox{\boldmath${I}$}$ is given by a combination
    \begin{eqnarray}
    &&{\mbox{\boldmath$I$}}=
    2\,\varepsilon_{\cal D}\,
    (\,\varPsi_+\varPsi'_+
    -\varPsi_-\varPsi'_-\,),\,\,\,\,\,\varepsilon_{\cal D}
    =\frac{\cal D}{|\,{\cal D}\,|}=\pm 1,                 \label{bfI}\\
    &&\varPsi_\pm\equiv
    \varPsi(\tau_\pm),\,\,\,\,
    \varPsi'_\pm\equiv\varPsi'(\tau_\pm),    \label{Psis}
    \end{eqnarray}
of boundary values at these points of a special (non-periodic)
solution $\varPsi(\tau)$ of the homogeneous equation
${\mbox{\boldmath${F}$}}\,\varPsi(\tau)=0$ with the same operator
(\ref{operator}). This solution is defined by the equation
    \begin{eqnarray}
    \varPsi(\tau)\equiv
    g(\tau)\int_{\tau_*}^{\tau}\frac{dy}{g^2(y)},\,\,\,\,\,
    \tau_-\equiv 0<\tau<\tau_+,\,\,\,\,\,
    \tau_-<\tau_*<\tau_+,                              \label{Psi}
    \end{eqnarray}
where $\tau_*$ is some fixed point belonging to this half period
range of $\tau$.

The main property of the function $\varPsi(\tau)$ is that it is
smoothly defined in the half-period range of $\tau$ and $\tau_*$,
where the integral (\ref{Psi}) is convergent because the roots of
$g(\tau)$ do not occur in the integration range of (\ref{Psi}). It
cannot be smoothly continued beyond this half-period, though its
limits are well defined for $\tau\to\tau_\pm\mp0$,
    \begin{eqnarray}
    \varPsi(\tau_\pm)=
    -\frac1{g'(\tau_\pm)}\equiv
    -\frac1{g'_\pm},                  \label{Psipm}
    \end{eqnarray}
because the factor $g(\tau)$ tending to zero compensates for the
divergence of the integral at $\tau\to\tau_\pm$. Moreover, because
of $g''(\tau_\pm)=0$ the function $\varPsi(\tau)$ is differentiable
at $\tau\to\tau_\pm$, and all the quantities which enter the
algorithm (\ref{bfI}) are well defined. These properties of
$\varPsi(\tau)=\varPsi(\tau,\tau_*)$ guarantee that
$\mbox{\boldmath$I$}$ is independent of an arbitrary choice of the
point $\tau_*$, which can be easily verified by using a simple
relation $d\varPsi'_\pm/d\tau_*=-g'_\pm/g^2(\tau_*)$.

The structure of (\ref{prefactor0}) suggests that the effect of the
one-loop prefactor can be strong, because the argument of the square
root can tend to zero. Though the factor $d^2F/d\eta^2$ which is
proportional to the negative of the specific heat of the thermal
bath,
    \begin{eqnarray}
    \frac{d^2F}{d\eta^2}=
    -\sum_\omega\frac{\omega^2\,e^{\omega\eta}
    }{\big(e^{\omega\eta}\mp 1\big)^2}<0,     \label{specificheat}
    \end{eqnarray}
is always negative, the sign of ${\mbox{\boldmath$I$}}$ is
indefinite. Indeed, this quantity can be rewritten as the action
functional of the mode $\varPsi(\tau)$ in the half period of the
Euclidean time range
    \begin{eqnarray}
    {\mbox{\boldmath$I$}}=2\,\varepsilon_{\cal D}
    \int_{\tau_-}^{\tau_+}
    d\tau\,\left(\,\varPsi'^2+\frac{g''}g\,\varPsi^2\right).
    \end{eqnarray}
Its sign depends on the sign of the Hessian matrix in (\ref{bfI})
(ghost or non-ghost nature of scale factor perturbation $\varphi$ in
the gravitational action). Moreover, the integral here is also
indefinite, because the potential term is not positive-definite. In
fact it is basically negative ($\varphi$ is a tachyon), because for
an oscillating $g(\tau)\sim a'(\tau)$ this function is
quasi-harmonic with $g''/g<0$ in the average over the oscillation
period. In particular, for a constant $g''/g=-w^2$, the function
$g(\tau)$ is harmonically oscillating, the function $\varPsi(\tau)$
is exactly calculable and ${\mbox{\boldmath$I$}}$ identically
vanishes -- the case of the second zero mode of
${\mbox{\boldmath$F$}}$ complimentary to $g(\tau)$ \cite{det},
    \begin{eqnarray}
    &&g(\tau)=d\,\sin(w\tau),\,\,\,\,
    \varPsi(\tau)=
    \frac{\sin\big(w(\tau-\tau_*)\big)}
    {d\,\sin(w\tau_*)},\,\,\,\,
    \varPsi_+\varPsi'_+
    -\varPsi_-\varPsi'_-=0.           \label{harmonic}
    \end{eqnarray}
Therefore, a nontrivial expression for the prefactor is based on the
anharmonicity of $g(\tau)$ and should be analyzed numerically or by
some kind of perturbation theory.

\section{Estimates for the prefactor in the CFT driven cosmology}
Now we apply the above technique to the model of the CFT driven
cosmology described in Introduction. As was shown in \cite{slih},
all saddle-point instantons of this model exist only inside the
curvilinear wedge in the two-dimensional plane of the variables
$(H^2,C)$ -- primordial cosmological constant $\Lambda=3H^2$ and the
amount of radiation constant $C$, depicted in Fig.1. This wedge is
bounded by the lower straight-line and upper hyperbolic boundaries,
    \begin{eqnarray}
    B-B^2H^2<C<1/4H^2.     \label{wedge}
    \end{eqnarray}
The one-parameter families of k-fold instantons form in this
plane the curves interpolating between these two boundaries, and
their sequence accumulates at the critical point corresponding to
the quantum gravity scale $H^2_{\rm max}=1/2B$. Each instanton
solution is represented by a point on one of these curves with the
corresponding $k$ -- the number of oscillations of the scale factor
$a$ between its minimal and maximal values. The range of $\Lambda$ in which these $k$-fold instantons exist is given by a projection of their family on the axes of the variable $H^2$. Here we consider only the one-fold family.

\begin{figure}[h]
\centerline{\epsfxsize 12cm \epsfbox{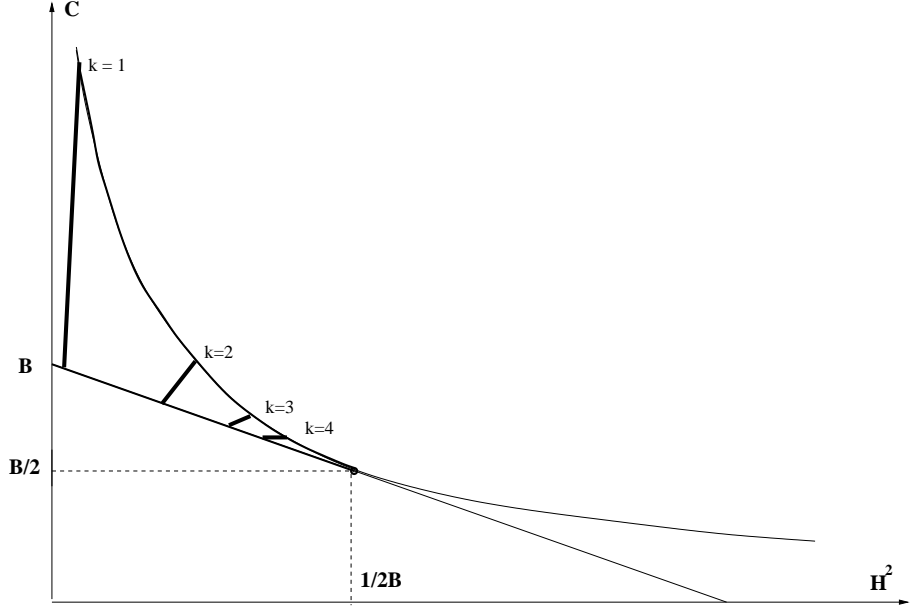}} \caption{\small
Instanton domain in the $(H^2,C)$-plane. Instanton families are
shown for $k=1,2,3,4$. Their sequence accumulates at the critical
point $(1/2B,B/2)$.
 \label{Fig.1}}
\end{figure}

Analytic estimates of the prefactor are hard to make even in the
limit of a large number of quantum fields $\mathbb{N}\gg 1$ and a
large value of $B$, because as we will now see this one-loop factor
is $O(1)$ and is not sensitive to this limit. A natural quantity
that could have played the role of a smallness parameter is the
combination
    \begin{eqnarray}
    \varepsilon=1-2BH^2,     \label{varepsilon}
    \end{eqnarray}
which belongs to the bounded range $0\leq\varepsilon\leq1$ for all
these instantons. As was shown in \cite{slih} for multi-fold
instantons with $k\gg 1$ this quantity is indeed small
$\varepsilon\simeq\ln k^2/2\pi^2k^2\to 0$ and corresponds to
solutions tending to the corner of the curvilinear triangle $C=B/2$
-- the point of a new quantum gravity scale $H^2=1/2B$.
Unfortunately, for the case of $k=1$ instanton which only we
consider here $\varepsilon=O(1)$ and the corresponding prefactor
should be calculated numerically.

To make this calculation manageable let us first parameterize the
instanton solutions in a more convenient way in terms of the
quantity (\ref{varepsilon}). From the modified Friedmann equation
(\ref{efeq}) we know that the scale factor is a function of time
oscillating between the two turning points $a_\pm$. In terms of a
dimensionless variable
    \begin{eqnarray}
    z=\frac{a^2}B      \label{z}
    \end{eqnarray}
this behavior can be expressed as a solution of (\ref{efeq}) for the
time derivative of $a(\tau)$
    \begin{eqnarray}
    &&a'^2=(1-\varepsilon)\frac{(z_+-z)(z-z_-)}
    {\sqrt{\varepsilon\,\big(z^2-z_c^2\big)}+z-1},
    \end{eqnarray}
where the turning points $z_\pm$ and $z_c=a_c^2/B$ read in terms of
$B$, $C$ and $\varepsilon$ as
    \begin{eqnarray}
    &&z_\pm=\frac1{1-\varepsilon}
    \left(1\pm\sqrt{1-\frac{2C}B
    (1-\varepsilon)}\,\right),       \\
    &&z_c^2=\frac{a^4_c}{B^2}
    =\frac1\varepsilon\left(\frac{2C}B-1\right).
    \end{eqnarray}

In terms of $\varepsilon$ the curvilinear wedge (\ref{wedge}) where
the instantons are located looks as
    \begin{eqnarray}
    \varepsilon<\frac{2C}B-1
    <\varepsilon+\frac{\varepsilon^2}{1-\varepsilon}.
    \end{eqnarray}
Inside this domain the one-parameter family of solutions
interpolating between its boundaries can be labeled by the parameter
$d$ ranging from $d=0$ at the upper boundary to $d=1$ at the lower
boundary
    \begin{eqnarray}
    \frac{2C}B-1=\varepsilon+(1-d^2)\,
    \frac{\varepsilon^2}{1-\varepsilon},
    \,\,\,\,0\leq d\leq 1.                  \label{Cvsepsilon}
    \end{eqnarray}
Together with the definition (\ref{varepsilon}) this relation is
just a relabeling  of the solutions from the dimensional variables
$(C,H^2)$ to the dimensionless $(\varepsilon,d)$ having a simple
finite range inside a unit quadrangle.

In terms of $d$ the turning points of the solution $z_\pm$ and the
parameter $z_c$ introduced above take the form
    \begin{eqnarray}
    z_\pm=\frac{1\pm\varepsilon\,d}{1-\varepsilon},\,\,\,\,
    z_c^2=\frac{1-\varepsilon\, d^2}{1-\varepsilon}
    \end{eqnarray}
With $z$ parameterized between the two turning points by the new
variable $x$
    \begin{eqnarray}
    z=z_-+\frac{2\varepsilon\, d}{1-\varepsilon}\,x,
    \,\,\,\,0\leq x\leq 1                  \label{x}
    \end{eqnarray}
the equation for $a'$ can be written down as
    \begin{eqnarray}
    &&a'^2=4\varepsilon d^2\,\frac{x(1-x)}
    {\big[\,(1-d)^2+4d\,(1-\varepsilon\,d)\,x
    +4\varepsilon\,d^2 x^2\,\big]^{1/2}+1-d+2d\,x},
    \end{eqnarray}
so that the conformal time period of the instanton equals
    \begin{eqnarray}
    &&\eta=2\int\limits_{a_-}^{a_+}
    \frac{da}{a'a}=\sqrt\varepsilon
    \int\limits_0^1 \frac{dx}{x^{1/2}(1-x)^{1/2}}\nonumber\\
    &&\qquad\qquad\qquad\times\,\frac{\left(\,[\,(1-d)^2
    +4d\,(1-\varepsilon\,d)\,x
    +4\varepsilon\,d^2 x^2\,]^{1/2}
    +1-d+2d\,x\right)^{1/2}}
    {(1-\varepsilon d+2\varepsilon^2 d x)}.   \label{eta10}
    \end{eqnarray}
With a very good accuracy for all $\varepsilon$ and $d$ the integral
here equals $\pi\sqrt2$, so that $\eta\simeq\pi\sqrt{2\varepsilon}$.
On the other hand, the bootstrap equation (\ref{bootstrap}) for
$\varepsilon$ in view of (\ref{Cvsepsilon}) reads
    \begin{eqnarray}
    \frac{2C}B-1\equiv\varepsilon+(1-d^2)\,
    \frac{\varepsilon^2}{1-\varepsilon}
    =\frac2{m_P^2B}\frac{dF}{d\eta}.      \label{bootstrap1}
    \end{eqnarray}
In view of the fact that both $B\sim\mathbb{N}$ and
$dF/d\eta\sim\mathbb{N}$ for a large number of species $\mathbb{N}$,
the ratio on the right hand side here is $O(\mathbb{N}^0)=O(1)$, and
the parameter $\varepsilon$ also remains $O(1)$ and is not sensitive
to the limit of large $\mathbb{N}$. Then one can easily check that
in the limits of either small, $\eta\ll 1$, $dF/d\eta\sim1/\eta^4$,
or relatively large, $\eta\sim\pi$, $dF/d\eta\sim\exp(-\#\eta)$,
values of $\eta$ this equation yields the value of $\varepsilon$
close to the middle of its admissible range $0\leq\varepsilon\leq
1$, $\varepsilon\sim 1/2$. This confirms that for a one-fold garland
instanton $\varepsilon$ never gets either very small or close to
one, and numerical simulations become inevitable for the calculation
of the prefactor (\ref{prefactor0}) with its basic quantity
(\ref{bfI}). Fortunately, this leads to a simple qualitative
picture.

The function (\ref{g}) for the Lagrangian of (\ref{effaction})
equals in terms of the variable $z$, (\ref{z}), and the running
integration parameter $x$ of Eq.(\ref{x})
    \begin{eqnarray}
    &&g^2=12\pi^2\,M_P^2B\,a\,a'^2\,
    \sqrt{\varepsilon\,\big(z^2-z_c^2\big)},   \label{g1}\\
    &&\sqrt{z^2-z_c^2\vphantom{L^L}}
    =\frac{\sqrt\varepsilon}{1-\varepsilon}\,
    \big[\,\big(1-d\big)^2+4d\big(1-\varepsilon\,d\big)\,x
    +4\varepsilon\,d^2 x^2\,\big]^{1/2}.
    \end{eqnarray}
Therefore the function (\ref{Psi}) is given by the following
integral
    \begin{eqnarray}
    &&\varPsi(\tau)=g(\tau)\int\limits_{a_*}^{a(\tau)}
    \frac{da}{a'\, g^2}=\frac{g(\tau)}
    {16\,m_P^2\,B}\,\frac1{d^2}
    \int\limits_{x_*}^{x(\tau)}
    \frac{dy}{y^{3/2}\,(1-y)^{3/2}}\,G(y),    \label{Gint}\\
    &&G(y)=\frac1{\varepsilon^{3/2}}\,
    \left(\frac{1-\varepsilon}
    {1-\varepsilon d+2\varepsilon^2 d y}\right)^{3/2}\nonumber\\
    &&\qquad\qquad\qquad\times\,\frac{\left(\,[\,(1-d)^2
    +4d\,(1-\varepsilon\,d)\,y
    +4\varepsilon\,d^2 y^2\,]^{1/2}
    +1-d+2d\,y\right)^{3/2}}
    {\big[\,\big(1-d\big)^2+4d\big(1-\varepsilon\,d\big)\,y
    +4\varepsilon\,d^2 y^2\,\big]^{1/2}},
    \end{eqnarray}
where the integration limits equal
    \begin{eqnarray}
    &&x(\tau)=\frac{1-\varepsilon}{2\varepsilon d}\,
    \frac{a^2(\tau)-a_-^2}{\varepsilon\,B},\,\,\,\,
    x_*=\frac{1-\varepsilon}{2\varepsilon d}\,
    \frac{a^2_*-a_-^2}{\varepsilon\,B}.
    \end{eqnarray}

In the limit $\tau\to\tau_\pm$ (equivalent to $x(\tau)\to 1$ and
$x(\tau)\to 0$) the integral in (\ref{Gint}) is divergent, this
divergence being compensated by the factor $g(\tau)$ vanishing at
these points. For the calculation of (\ref{bfI}) we have to take
these limits, and their calculation is based on a preliminary
integration by parts in $y$ which disentangles the divergent part of
the integral, $\sim 1/\sqrt{1-x(\tau)}$ at $x(\tau)\to 1$ or $\sim
1/\sqrt{x(\tau)}$ at $x(\tau)\to 0$. Then multiplying the result by
$g(\tau)$ and differentiating we get in view of (\ref{Psipm}) (we
take into account that for the CFT driven cosmology
$\varepsilon_{\cal D}=-1$)
    \begin{eqnarray}
    &&\mbox{\boldmath$I$}=-2(\varPsi_+\varPsi'_+
    -\varPsi_-\varPsi'_-)
    =-\frac{G}{8\,d^2 m_P^2
    B},                          \label{bfIfinal}\\
    &&G=\int\limits_{x_*}^1
    dy\,\frac{G(y)-G(1)\,y^{3/2}}{y^{3/2}(1-y)^{3/2}}\nonumber\\
    &&\quad\qquad\quad+
    \int\limits_0^{x_*}
    dy\,\frac{G(y)-G(0)\,(1-y)^{3/2}}{y^{3/2}(1-y)^{3/2}}
    -\frac{2\,G(1)}{(1-x_*)^{1/2}}-
    \frac{2\,G(0)}{{x_*}^{1/2}}.               \label{G}
    \end{eqnarray}
This expression is, of course, independent of the choice of the
intermediate point $x_*$, as is easily seen by differentiation with
respect to $x_*$.

For $d\to 0$ the function $G=O(d^2)$, so that $\mbox{\boldmath$I$}$
is finite in this limit even despite $d^2$ in the denominator of
(\ref{bfIfinal}). Note that the limit $d\to 0$ corresponds to the
upper hyperbolic boundary of the instanton domain (\ref{wedge}) on
which the cosmological model is static, $a(\tau)=a_\pm=1/H\sqrt2$,
and $g''/g=-2(1-\varepsilon)/B\,\varepsilon=-w^2$ is constant. So
this is the case of the harmonic function $g(\tau)$ in
Eq.(\ref{harmonic}) for which formally $\mbox{\boldmath$I$}=0$ --
the case of the second zero mode of the operator
$\mbox{\boldmath$F$}$, discussed in \cite{det}. However, in this
limit not only the deviation from harmonicity $O(d^2)$ disappears,
but also the amplitude of the function $g(\tau)\sim d$ tends to
zero, cf. Eq.(\ref{g1}). Therefore, in view of (\ref{harmonic}) the
function $\mbox{\boldmath$I$}(d)=O(d^2)/d^2$ acquires a $1/d^2$
factor, and does not vanish for $d\to 0$. Thus, quantum corrections
to the preexponential factor are nontrivial even in this limit. As
we will see below, they are smaller than those of $d\to 1$ but still
nonzero.

Numerical calculation of (\ref{G}) can be approximately done in the
full range of $d$. First we find from the bootstrap equation
(\ref{bootstrap1}) the value of $\varepsilon$ as a function of $d$
on the relevant one-parameter family of instantons $0\leq d\leq 1$, $\varepsilon=\varepsilon(d)$, and then evaluate with the aid of (\ref{bfIfinal}), (\ref{G}) and (\ref{specificheat}) the preexponential factor on this family
    \begin{eqnarray}
    P(d)=\left|\,1-\mbox{\boldmath$I$}\,
    \frac{d^2F}{d\eta^2}\,\right|^{-1/2}_{\;\;\;
    \varepsilon=\varepsilon(d)}.               \label{P(d)}
    \end{eqnarray}

We will explicitly demonstrate this procedure on the example of a single gauge vector field. For this field the trace anomaly coefficient $B$ and the sum (\ref{energy}) over its oscillator frequencies $\omega=\sqrt{2n^2-4}$, $n\geq 2$, read as
    \begin{eqnarray}
    m_P^2B=\frac{31}{120},\,\,\,\,
    \frac{dF}{d\eta}=2\sum\limits_{n=2}^\infty(n^2-1)\,
    \frac{\sqrt{2n^2-4}}
    {e^{\,\eta\,\sqrt{2n^2-4}}-1}\,.
    \end{eqnarray}
With these expressions and the conformal time $\eta$ given by the integral (\ref{eta10}) the numerical solution of the bootstrap equation (\ref{bootstrap1}) can be sufficiently accurately fitted by the following function
    \begin{eqnarray}
    \varepsilon_{\rm vector}(d)= 0.345-0.096\,d+0.042\,d^2,
    \end{eqnarray}
which parameterizes the one-fold family of instantons. Then we compute $\mbox{\boldmath$I$}$ via (\ref{bfIfinal})-(\ref{G}), which turns out to be negative on this family, and also evaluate the specific heat for the free energy of vector particles on a 3-sphere
    \begin{eqnarray}
    \frac{d^2F}{d\eta^2}=
    -2\sum\limits_{n=2}^\infty(n^2-1)\,\frac{(2n^2-4)\,
    e^{\,\eta\,\sqrt{2n^2-4}}}
    {\big(e^{\,\eta\,\sqrt{2n^2-4}}-1\big)^2}\,.
    \end{eqnarray}
The use of (\ref{P(d)}) for this family of solutions finally
yields the plot of $P(d)$ depicted on Fig.2. A characteristic feature of this vector field case is that with $d^2F/d\eta^2<0$ and negative
$\mbox{\boldmath$I$}$ the denominator of $P(d)$ is smaller than one,
but never degenerates to zero. Therefore, the one-loop prefactor
stays slightly higher than one, but remains finite and within the
perturbation theory range.
\begin{figure}[h]
\centerline{\epsfxsize 12cm \epsfbox{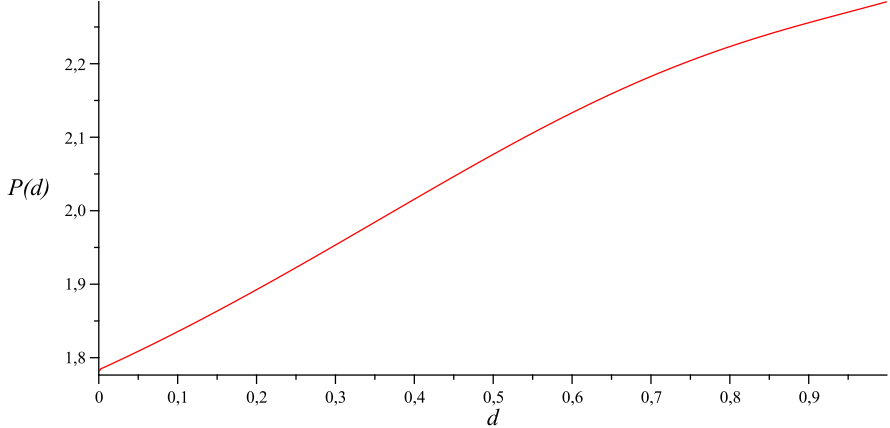}} \caption{\small
Plot of $P(d)$ for for a gauge vector field. \label{Fig.2}}
\end{figure}

Similar calculations can be done for pure massless spinor and
conformal scalar fields. The prefactor for these fields remains nearly constant throughout the whole family of one-fold instantons. For a spinor field the behavior of $P(d)\simeq 1.1>1$ is qualitatively the same as in the vector field case, whereas for a
scalar field the calculations show that $\mbox{\boldmath$I$}>0$ and  the prefactor turns out to be smaller than one, $P(d)\simeq 0.9<1$. Possible mixtures of various spins do not qualitatively change this picture.

\section{Conclusions}

The one-loop prefactor of the statistical sum in the CFT driven
cosmology was obtained for a limited set of saddle-point solutions
having one oscillation of the cosmological scale factor. This
limitation is currently explained by the fact that the major
ingredient of the calculational algorithm -- the restricted
functional determinant of the operator $\mbox{\boldmath$F$}$ is
known in a closed form (\ref{DetFstar}) only for this simplest set
of instantons \cite{det}. The resulting preexponetial factor does
not present us with new physics, because for all scalar, spinor and
vector conformal particles or their combinations the prefactor
remains a smooth function $P=O(1)$. It justifies the semiclassical
expansion, but does not feature phase transitions or strong coupling
range beyond perturbation theory restrictions even in the limit of
numerous conformal species $\mathbb{N}\gg1$. At the same time path
integral applications in such conformal cosmology suggest the
existence of instantons with an arbitrary number of oscillations of
the cosmological scale factor. In particular, for the number of
these oscillations tending to infinity this CFT driven cosmology
approaches a new quantum gravity scale $\Lambda_{\rm max}=3H^2_{\rm
max}=3/2B$ -- the maximum possible value of the cosmological
constant \cite{slih,why} -- where the physics and, in particular,
the effects of the quantum prefactor become very interesting and
important. Thus the extension of the above results to the full set
of cosmological instantons looks very promising, as this extension
might be relevant to the resolution of the cosmological constant and
landscape problems \cite{why}. We hope to attain this extension in
the foreseeable future.

Main difficulty with all known applications of quantum gravity and cosmology is that the lack of UV complete theory forces us to work within the effective field theory trustable only below its cutoff. In this respect large $\mathbb{N}$ CFT driven cosmology of \cite{slih} seems to be a safe theory because its instanton curvature scale is bounded by $H^2_{\rm max}=1/2B\sim m_P^2/\mathbb{N}\ll m_P^2$ which is much lower than the Planck scale. Therefore, no non-perturbative extension (like the hypothetical asymptotic safety of quantum gravity \cite{asymptoticsafety}) is likely to be needed, and a usual semiclassical expansion is sufficient to handle the problem of cosmological initial conditions. However, as persuasively advocated in \cite{species}, the cutoff in theories with a large number of quantum species, $\sim m_P^2/\mathbb{N}$, also decreases with the growing $\mathbb{N}$ and the instantons considered above turn out to be very close to this cutoff scale. This makes important the computation of loop corrections initiated above. Finiteness of the one-loop order and its regularity $P=O(1)$ in the full one-fold instanton range of $\Lambda$ gives a hope that this property will survive beyond this range and beyond one-loop approximation, and this is a subject of our further studies \cite{progress}.

\section*{Acknowledgements}
I am grateful to A.Yu.Kamenshchik for helpful discussions. Also I
wish to express my gratitude to G.Dvali for hospitality at the
Physics Department of the Ludwig-Maximilians University in Munich
where this work was supported in part by the Humboldt Foundation.
This work was also supported by the RFBR grant No 11-02-00512.

\end{document}